\documentclass[10pt]{article}
\usepackage{array,amsmath,amsfonts,graphicx,mathrsfs}
\usepackage[colorlinks=true]{hyperref}

\newcommand{\bs}{\boldsymbol}

\newtheorem{theorem}{Theorem}

\begin{document}

\centerline{\Large\bf On the Role of Decision Theory}
\centerline{\Large\bf in Uncertainty Analysis}

\vspace{2pt}

\vspace{.4cm}
\centerline{Merlin Keller$^{1,2},$ Eric Parent$^2$, Alberto Pasanisi$^{1},$}
\vspace{.4cm}
\centerline{\it$^1$ EDF R\&D, Chatou, France}
\centerline{\it $^2$ AgroParisTech, Paris, France}

\vspace{.55cm}

\begin{quotation}

{\bf Abstract.} Maximum likelihood estimation (MLE) and heuristic predictive estimation (HPE) are two widely used approaches in industrial uncertainty analysis. We review them from the point of view of decision theory, using Bayesian inference as a gold standard for comparison. The main drawback of MLE is that it may fail to properly account for the uncertainty on the physical process generating the data, especially when only a small amount of data are available. HPE offers an improvement in that it takes this uncertainty into account. However, we show that this approach is actually equivalent to Bayes estimation for a particular cost function that is not explicitly chosen by the decision maker. This may produce results that are suboptimal from a decisional perspective. These results plead for a systematic use of Bayes estimators based on carefully defined cost functions.

\vspace{9pt}
 \par

\end{quotation}
\newpage

\tableofcontents
\newpage

\section{Introduction}\label{sec:introduction}

More and more applied industrial studies involve nowadays explicit uncertainty assessment \cite{Helton94,Pate-cornell96,deRocquigny08}. Indeed, in industrial practice, engineers coping with quantitative predictions using computer models must face numerous and quite different sources of uncertainty, which affect the results of their studies. In most practical problems, the objective is to study the probability distribution of the output of a deterministic model, the inputs of which are random variables.

Uncertainty on the model, due for instance to the prohibitive computational cost its implementation may require, or to its discrepancy with the physical system it represents, can be treated in this framework, which is sometimes called \textit{Uncertainty Analysis} \cite{Kennedy01}. However, for simplicity we do not tackle such sources of uncertainties in what follows. Furthermore, the model is assumed to provide deterministic outputs for each given set of inputs, that is, to each input value corresponds a unique output value.

Within this paper, we will voluntarily stay within this framework, which is rather common in the industrial practice \cite{deRocquigny08}. The ingredients of the problem are then:
\begin{itemize}
\item[-] a preexisting physical or industrial system, represented by a deterministic model $G(\cdot);$
\item[-] inputs of the physical model, affected by various sources of uncertainty, and modeled jointly as a random vector, noted $\rm X$ in the following. Hence we adopt probability theory as a means of quantifying uncertainties, and do not consider here the use of other tools, such as fuzzy logic \cite{Bardossy95,Zadeh05}, Dempster-Shafer Theory \cite{Shafer76,Walley91,Caselton92}, etc.
\item[-] outputs of the physical model, {\em i.e.}, deterministic functions of the inputs, which we note in the following: ${\rm Y} = G({\rm X}).$ In the following, we will limit ourselves for simplicity to scalar outputs, though our arguments can be extended to multivariate outputs;
\item[-] an amount of data and expertise available to assess input uncertainty,
\item[-] industrial stakes that motivate the uncertainty assessment more or less explicitly (safety and security, environmental control, resources optimization etc.).
\end{itemize}

The industrial stakes guide the choice of one or more particular \textit{quantities of interest} summarizing the main results of the uncertainty analysis, e.g. the mean and/or a given quantile of the model output.

Note that this framework also includes the case where the deterministic model $G(\cdot)$ is the identity function, i.e. the classical problem of predicting particular features of the probability distribution of an observed random variable.

Whatever the complexity of the pre-existing model (the identity function or a time-consuming computer model), apart from computational issues, the problem is, at least conceptually, the same: given some information about an observable variable, predict some quantities of interest either of its probability distribution or of a deterministic function of it. The problem is then statistical in essence and involves the two classical steps of statistical analysis: inference (induction) and prediction (deduction) \cite{Girard04}.

In the uncertainty analysis framework, the engineer takes (or not) explicitly into account the uncertainty affecting the estimation of the parameters of the distribution function of model inputs, by adding (or not) an \textit{additional} uncertainty layer. In some cases, he/she often ignores this uncertainty and simply considers that the chosen inference method, such as maximum likelihood estimation (MLE), provides the exact value of the parameters. Although other estimation techniques such as the moments or weighted moments, pseudo-likelihood, etc. can be considered, this paper emphasizes MLE as it is by far the most popular. The propagation of the probability distribution of the inputs through the deterministic model leads thus to a distribution of outcomes, as if the true value of the parameter were the MLE.

In some other cases, the uncertainty affecting the value of the parameters is modeled by using a proper probability distribution function. If the statistical analysis has been done in the Bayesian framework, this distribution function is the parameter's posterior distribution; otherwise the distribution of a particular estimator (most of the time the normal asymptotic approximation of the MLE) can be used \cite{deRocquigny08}.

Once this additional uncertainty level has been added, the resulting mixture probability distribution function (pdf) for $\rm X$ is propagated through the deterministic model, for instance using the double monte-Carlo approach. This yields a distribution of the model output, which is known as the {\em posterior predictive distribution} if the chosen setting is Bayesian. It is tempting to conclude \cite{Pasanisi09} that this predictive distribution summarizes all the sources of uncertainty affecting this output, and to evaluate the quantity of interest directly using this predictive distribution. For instance, if one is interested in a given quantile of the output, then, according to a common practice, one could often evaluate it as the corresponding quantile of the predictive distribution.

This heuristic approach, in which essentially the inference problem is separated from the prediction phase of the statistical analysis, can lead to erroneous results as it is not coherent with statistical decision theory, as we will show in the following. The main purpose of this paper is to put standard engineering practices into a more formal framework so as to point out the (more or less) hidden hypotheses underlying the heuristic approach. Avoiding any dogmatic temptation, this paper aims to let the uncertainty practitioner be aware of these underlying hypotheses, to let he/she decide if they could be accepted or not, given the particular context of the industrial study.

The rest of this paper is organized as follows. We start in Section~\ref{sec:decision}, by recalling the basic principles of decision theory and Bayes estimation. This approach, known to be optimal from a decisional viewpoint, constitutes a gold standard to which other approaches may be compared. In Section~\ref{sec:MLE}, we discuss the relevance of MLE, whose justification relies on large sample properties, in industrial risk problems where the data are often scarce. The heuristic use of the predictive distribution briefly sketched above is reviewed in Section~\ref{sec:predictive}. We show that the use of this seemingly generic heuristic is equivalent to Bayes estimation, and induces implicitly the choice of a particular cost function that depends on the mathematical expression of the chosen quantity. Section~\ref{sec:dyke} compares all reviewed approaches on a dyke reliability assessment problem toy example. These results and their implications for industrial uncertainty analysis are discussed in Section~\ref{sec:discussion}.

\section{Decision theory and Bayes estimators}\label{sec:decision}

\subsection{General problem formulation}\label{sec:formulation}

In the following, the model inputs will be modeled jointly as a random vector ${\rm X},$ distributed according to 
an unknown probability distribution function $F.$ However, instead of dealing with such a functional uncertainty, we restrict ourselves to look for a distribution belonging to a parametric family, admitting a pdf $f(\cdot | \theta),$ specified by an unknown parameter vector $\theta.$

The model output is noted ${\rm Y} = G({\rm X}),$ and its pdf is noted $p(\cdot | \theta)$ in the following. We additionally assume that the physical process represented by $G$ is not random. When $G$ is invertible, the output's pdf is thus given by the formula: $p(y | \theta) = f(G^{-1}(y) | \theta) / | G' (G^{-1}(y)) |.$

We also assume, as is in general the case, that the data are a sample $D = (x_1, \ldots, x_n)$ of input values, that is, independent realizations of the variable ${\rm X}.$ We note $\mathcal L(D|\theta) = \Pi_{i=1}^n f(x_i| \theta)$ the likelihood of the data vector. Based on this model and the data $D,$ our goal is twofold:

\begin{itemize}
\item[{\em i.}] Infer the law of $\rm X,$ {\em i.e.} in our case estimate $\theta;$
\item[{\em ii.}] Deduce an estimate of the pdf of $\rm Y,$ and of a certain quantity of interest (mean, quantile, etc.) of this law, that is, a certain function $\phi = \phi(F)$ of the inputs' distribution. In the parametric framework we have adopted, this reduces to a function $\phi(\theta)$ of the model parameters.
\end{itemize}

\paragraph{Illustrative example}

Imagine we wish to assess the lifetime $\rm X$ of an industrial component, based on the previously recorded lifetimes $x_1, \ldots, x_n$ of identical components, generated by the same production chain. As is a classical approach in survival analysis, we postulate that $D = (x_1, \ldots, x_n)$ is a sample from the exponential distribution $\mathcal E(\theta),$ with pdf $f(x | \theta) = (1 / \theta) e^{-x/\theta}.$ Here the parameter $\theta$ is seen to be the average lifetime, that is, $\mathbb E[{\rm X} | \theta] = \theta.$ Note that we have no physical model here, so that $G(\cdot)$ reduces to identity, and $\rm Y = X.$

Once this model has been established, we can compute quantities of interest, for instance the {\em reliability} of the component, that is, the probability that it lasts more than $t$ units of time, which we can write as: $\phi = \mathbb P[{\rm X} > t | \theta] = e^{-t / \theta}.$ If the component under study is a light bulb for instance, $t = 2\ {\rm years}$ corresponds to standard regulations in order to ensure some acceptable working life period.

In practice, $\phi,$ like $\theta,$ is unknown but we can use the data $D$ to estimate this quantity of interest by $\hat \phi = \hat \phi(x_1, \ldots, x_n).$ If this estimated reliability is judged too small, then the production chain might be stopped, or certain of the machines involved in the process may have to be fixed or changed, etc.

For example, if $\hat \phi < 1\%$ then the production will be discarded, if $1\% \leq \hat \phi < 15\%$ then the production will be sold at a lower price with reduced guarantee; if $\hat \phi \geq 15\%$ then the production will be released at the ordinary price with the usual guarantee.
\newline

In other words, the mapping:
\begin{eqnarray}
(x_1, \ldots, x_n) &\mapsto& \hat \phi = \hat \phi(x_1, \ldots, x_n)\nonumber
\end{eqnarray}
allows for a {\em decision rule} between what is known from the piece of information carried by the sample $D$ and the possible final decisions that the industry manager can make, such as: $\{$discard, sell at a lower price, release$\}$ in the previous example. By extension, we will call a possible value $d$ for $\hat \phi$ a `decision', even though the line of reasoning should go one step further to reach the operational decision.

\subsection{Cost function}\label{sec:cost}

As illustrated above, in most industrial studies, the estimated value $\hat \phi$ of the quantity of interest is used to take a certain decision. Because each decision has a context-dependent cost, intuitively the aim of the predictioner is to choose the value of $\phi$ which minimizes the anticipated value of the cost.

This notion of prediction is formalized in the context of decision theory by the concept of {\em loss} or {\em cost function}, that is, a certain function $C(\phi, d)$ that measures the cost resulting from decision $d$ when the quantity of interest is $\phi.$ Here the `decision' $d$ is simply the chosen value for the unknown quantity $\phi.$

To continue with the previous example, imagine that the true $\phi$ is $2.1\%$ and, due to the imperfect partial information we choose $d = 75\%.$ In other words we did not see that the situation was out of control ($\phi = 2.1\%$) and keep business as usual ($d = 75\%$ therefore release the production in ordinary conditions). This error may damage the company's reputation and lead to the loss of market shares, quantified by the cost $C(\phi=2.1\%, d = 75\%).$ This cost may be substantially higher than the cost $C(\phi=60\%, d = 10\%),$ associated to a false alarm situation in the light bulb industry.

However, the precise expression of the cost function is not always available to the engineer in charge of estimating $\phi,$ either because it has not been provided explicitly by the decision maker, or because the exact costs are hard to evaluate. In such cases, a reasonable expression for $C(\phi, d)$ must be chosen, which shares the main features of the real cost function, according to the available information.

It is usually required that this `surrogate' cost function respects some reasonable properties. For instance, the cost is usually chosen to be convex and minimal when $\phi$ is matched by the decision $d:$ $\min_d C(\phi, d) = C(\phi, \phi)$ \cite{Ulmo73}.

A useful concept, closely related to that of cost function, is the concept of {\em regret function}, also termed {\em opportunity loss} \cite{Howard89}. It is defined as the difference between the cost of decision $d$ and the cost of the optimal decision $\phi:$
\begin{eqnarray}
R_e(\phi, d) &=& C(\phi, d) - C(\phi, \phi).\nonumber
\end{eqnarray}
In other terms, $R_e(\phi, d)$ is the additional cost one has to pay when taking a decision $d$ different from the optimum. It can also be interpreted as a measure of information value, since it is the maximum price the decision maker is willing to pay to learn the optimal decision $\phi$ \cite{Bernier81}. Arguably, losses more naturally describe real-life problems than cost functions, and are therefore easier to define, when the true costs are not known precisely, as illustrated in the following examples.

\paragraph{Usual loss functions}\label{sec:costs}

\begin{figure}
\centering
\includegraphics[width=0.8\textwidth]{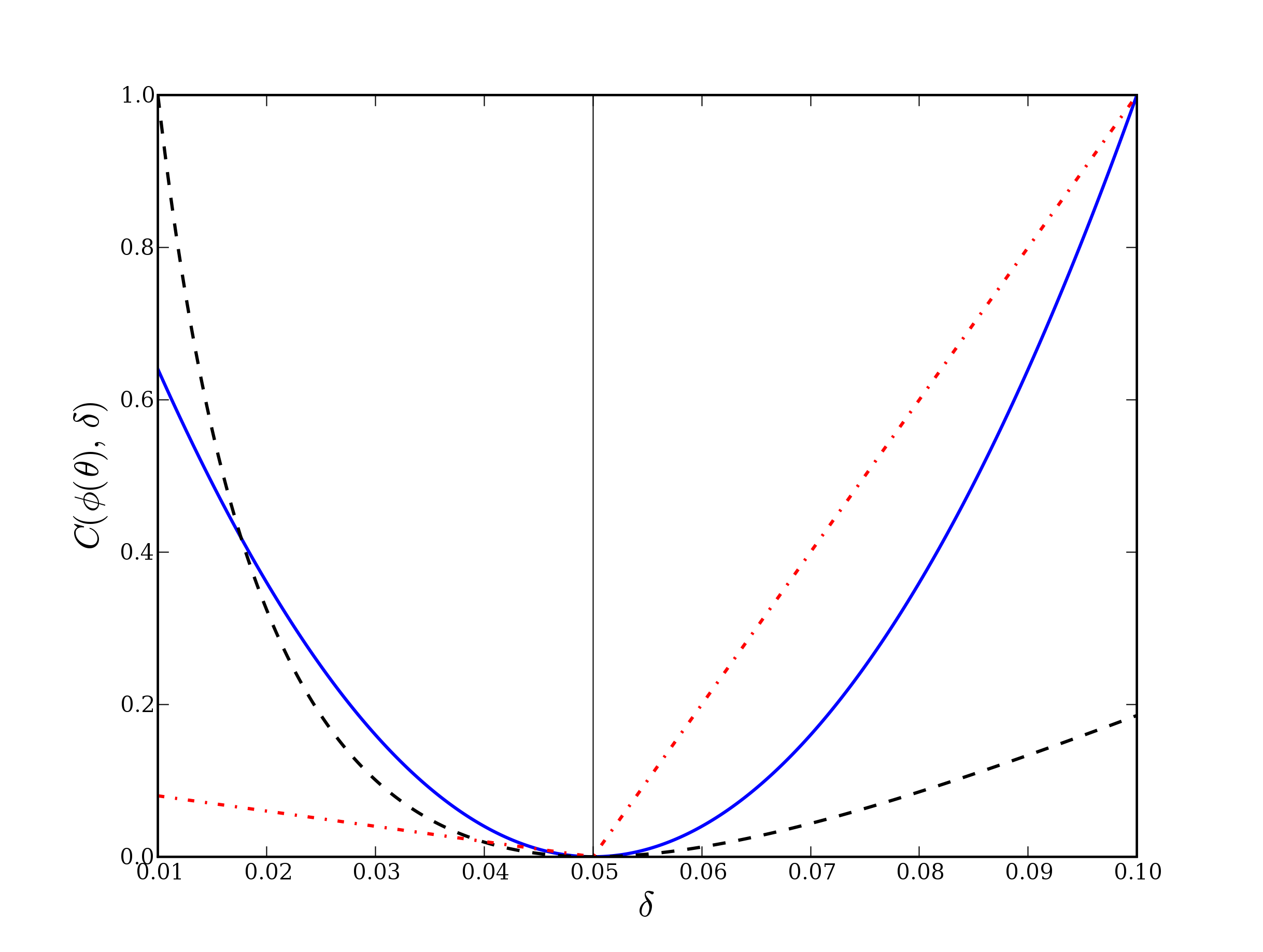}
\caption{\label{fig:cost_functions} \small Cost functions for the estimation of a tail probability. Continuous line: quadratic loss, dashed line: log-quadratic loss, dash-dotted line: weighted absolute loss, with $C2 = 10 \times C1.$ The vertical line corresponds to the true value $\phi.$ Cost functions are normalized for viewing convenience.}
\end{figure}

A popular choice of cost function is the quadratic loss, defined by
\begin{eqnarray}\label{eq:quadratic_loss}
C(\phi, d) &=& C_0 \times (\phi - d)^2.
\end{eqnarray}
Its use implies that the cost is quadratic in the error on $\phi,$ and does not depend on its sign. It can be derived as a second-order approximation of the `true' cost function, $C_0$ representing a constant marginal cost. Note that according to this justification, one should also include the minimal cost term $C_m,$ corresponding to the optimal decision $\phi.$ Hence (\ref{eq:quadratic_loss}) is in fact more readily interpretable as a {\em regret} than truly a {\em cost} function.

Often used as a default choice, the quadratic loss is however inappropriate if the cost depends on whether $\phi$ is over- or under-estimated. For instance, as illustrated above, the failure of an industrial component may have much more costly consequences than those resulting from stopping or fixing the production chain. In this case an asymmetric cost function might be preferable, such as the weighted absolute loss:
\begin{eqnarray}\label{eq:absolute_loss}
C(\phi, d) &=& |\phi - d| \times \left( C_1 \bs 1_{\{d < \phi\}}  + C_2 \bs 1_{\{d > \phi\}} \right).
\end{eqnarray}
This loss signifies that the additional cost of decision $d$ is proportional to its absolute deviation from $\phi,$ multiplied by a different factor depending on wether $d$ over- or under-estimates $\phi.$ This loss function is interesting from an operational point of view, in that it allows the user to specify precisely what costs are involved for each type of error.

Though the above cost functions are the most common, many others have been used in decision-oriented estimation problems, such as the $\alpha-$absolute loss \cite{Ren04}, the LINEX \cite{Varian74} or the entropy loss \cite{Robert07}. Going back to the problem of tail probability estimation, another criticism that can be done about the use of the quadratic loss is that in general we are less interested in the probability itself than by its order of magnitude, that is, its logarithm. Thus, we might consider using the log-quadratic loss:
\begin{eqnarray}\label{eq:log_quadratic_loss}
C(\phi, d) &=& (\log \phi - \log d)^2.
\end{eqnarray}
All three losses are illustrated in Figure~\ref{fig:cost_functions}, in the case where $\phi$ is a tail probability. 

\subsection{Bayesian approach}\label{sec:Bayes}

\begin{figure}
\centering
\includegraphics[width=0.8\textwidth]{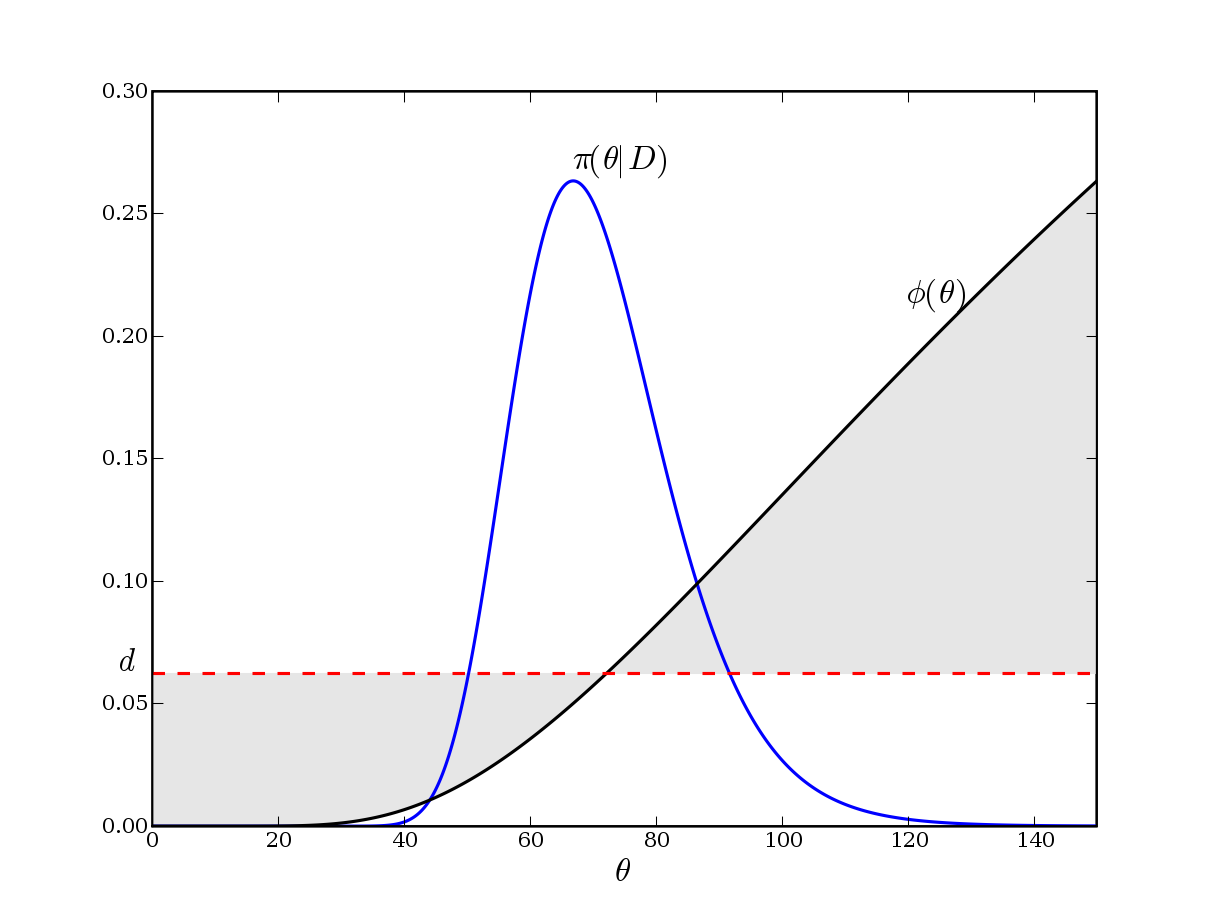}
\caption{\label{fig:sensitivity_analysis} \small Bayes estimation as an integrated sensitivity analysis. The Bayes estimate is the decision $d$ (dashed line) that minimizes the cost associated with the difference (shaded area) between the quantity of interest $\phi$ (thick line) and the decision $d,$ preferentially in regions of high posterior density $\pi(\theta | D)$ (thin line). In this particular example, $\phi$ is a tail probability in an exponential model, as explained in Section~\ref{sec:formulation}.}
\end{figure}

Minimizing the loss $C(\phi, d),$ or equivalently the regret $R_e(\phi, d),$ is impossible in practice because $\phi,$ like $\theta$ from which it is derived, is unknown (otherwise, in a perfect information situation, we would have the trivial solution $d = \phi$). Instead, the Bayesian approach consists in describing the uncertainty on the unknown parameter $\theta$ by a prior probability density $\pi(\theta),$ specifying the distribution of all possible values of $\theta.$ This prior information is updated using the data to yield the posterior distribution $\pi(\theta | D),$ according to Bayes' rule:

\begin{eqnarray}\label{eq:posterior}
\pi(\theta | D) &\propto& \mathcal L (D | \theta) \pi(\theta).
\end{eqnarray}
The optimal decision in this setting is obtained by minimizing the {\em expected posterior opportunity loss,} $\mathbb E [ R_e(\phi, d)| D],$ or equivalently, the {\em expected posterior loss}:
\begin{eqnarray}\label{eq:expected_loss}
\mathbb E [ C(\phi, d) | D] &=& \int_{\theta} C(\phi(\theta), d) \pi(\theta | D) d \theta.
\end{eqnarray}
As stressed in \cite{Parent07}, this procedure can be interpreted in terms of an {\em integrated sensitivity analysis}, where each possible cost resulting from decision $d$ is weighted according to the posterior probability of such a cost, evaluated conditionally on the available data. This is illustrated in Figure~\ref{fig:sensitivity_analysis}. Considered as a function of the sample $D,$ the decision that minimizes this quantity is called the Bayes estimator, relative to the prior~$\pi(\theta)$ and the loss $C(\cdot, \cdot).$ In the following, we will note this statistic: $\widehat{\phi}_{\textsc{bay}} = \widehat{\phi}_{\textsc{bay}}(D).$ 


Besides being intuitive, Bayes estimators are also known to have remarkable properties. Most importantly, they are systematically {\em admissible}, meaning that it is impossible to find an estimator with a uniformly lower risk \cite{Inoue09}, the risk of a decision rule $\delta$ (with $d = \delta(D)$) being defined as the average loss over all possible datasets, conditional on the true parameter value $\theta:$
\begin{eqnarray}\label{eq:risk}
\mathcal R(\phi(\theta), \delta) &:=& \int_{\tilde D} C(\phi(\theta), \delta(\tilde D)) \mathcal L (\tilde D | \theta) d \tilde D.
\end{eqnarray}
Thus, for a given estimation problem, once a cost function has been defined, it is in principle possible to consider only Bayes estimators as reasonable candidates, since all other rules are potentially {\em inadmissible}, under regularity conditions on the likelihood and the cost function \cite{Berger85}. Additionally, once a prior distribution on $\theta$ has been specified, a unique estimator is constructed pointwise ({\em i.e.} for each sample of information), by minimizing (\ref{eq:expected_loss}), which is equivalent to minimizing the Bayes risk,

\begin{eqnarray}\label{eq:expected_risk}
R_\pi(\delta) &:=& 
\int_{\theta} \mathcal R(\phi(\theta), \delta) \pi(\theta) d \theta,
\end{eqnarray}
over all the functional space of the decision rules $\delta$ \cite{Ulmo73}. This means that the choice of an optimal estimator, in the framework of decision theory, is uniquely determined by the problem specification (the cost function), and the amount of prior information available on the uncertain parameter.

\subsection{Bayes estimates for simple cost functions}\label{sec:examples}

In general, minimizing the posterior expected loss (\ref{eq:expected_loss}) may be difficult, depending on the choice of a loss function and prior distribution. However, the elementary cost functions mentioned earlier yield simple expressions for the corresponding Bayes estimate, which may explain partly their popularity.

\paragraph{Quadratic loss} It is straightforward to see that the Bayes estimate for the quadratic loss (\ref{eq:quadratic_loss}) is the posterior mean:
\begin{eqnarray}
\widehat{\phi}_{\textsc{bay}} &=& \mathbb E [\phi | D] \nonumber \\
&=& \int_{\theta} \phi(\theta) \pi(\theta | D) d \theta.\nonumber
\end{eqnarray}

\paragraph{Weighted absolute loss} Under the weighted absolute loss (\ref{eq:absolute_loss}), calculations are more involved. In \cite{Parent07}, it is shown that the resulting Bayes estimate $\widehat{\phi}_{\textsc{bay}}$ is such that:
\begin{eqnarray}
\mathbb P [ \phi < \widehat{\phi}_{\textsc{bay}} | D] &=& \frac{C_1}{C_1 + C_2}.\nonumber
\end{eqnarray}
In other terms, if $\frac{C_1}{C_1 + C_2} = 95 \%$ for instance, then $\widehat{\phi}_{\textsc{bay}}$ is the $95$-th percentile of the posterior density of the quantity of interest $\phi.$ In the special case where $C_1 = C_2,$ the optimal decision is seen to be the posterior median.

\section{MLE approach}\label{sec:MLE}

We now discuss the properties of the maximum likelihood estimation (MLE) approach, which is by far the most popular method of estimation currently used. The MLE of model parameters $\theta,$ noted $\widehat{\theta}_{\textsc{mle}}$ in the following, is simply the value from which the data are the most likely to have arisen:

\begin{eqnarray}
\widehat{\theta}_{\textsc{mle}} &:=& \arg \max_{\theta} \mathcal L (D | \theta).\nonumber
\end{eqnarray}
This quantity is then substituted to the unknown value of $\theta$ in the definition of $\phi,$ yielding the so-called `plug-in' estimate:
\begin{eqnarray}\label{eq:MLE}
\widehat{\phi}_{\textsc{mle}} &:=& \phi(\widehat{\theta}_{\textsc{mle}}).
\end{eqnarray}

A well-known property of the MLE is that it is invariant by reparameterization, so that if the model parameter $\theta$ is replaced by $\nu = \varphi(\theta),$ where $\varphi$ is one-to-one, then $\widehat \nu_\textsc{mle} = \varphi(\widehat \theta_\textsc{mle}),$ and the value of $\widehat{\phi}_{\textsc{mle}}$ remains unchanged. Note that this is not the case in general for Bayes estimates, which depend on the chosen parameterization.

\subsection{Asymptotic properties}\label{sec:asymptotic}

The MLE is engagingly simple compared to the Bayes estimators described in the previous sections, as it requires no specification of a cost function $C$ or a prior distribution $\pi.$ Furthermore, its use can be justified by its good asymptotic behavior (see \cite{Vandervaart00} for instance). First, the MLE is known to be {\em consistent} under very mild assumptions. This means that, for a sufficiently large number $n$ of observations, $\widehat{\theta}_{\textsc{mle}}$ becomes arbitrarily close to the true value $\theta.$ The same result thus applies to $\widehat{ \phi}_{\textsc{mle}},$ 
which tends to the true value $\phi$ when $n$ goes to infinity.

Secondly, under regularity conditions of the likelihood function $\mathcal L (D | \theta)$, the MLE is asymptotically normal, so that for $n$ `large enough' (in practice, $n = 40$ is considered sufficient in most applications), its law (conditional on the unknown value of $\theta$) can be approximated by:
\begin{eqnarray}\label{eq:asymptotic}
\mathscr L (\widehat{\phi}_{\textsc{mle}} | \theta) &\approx& \mathcal N \left(\phi(\theta), \frac{1}{n}\mathcal I^{-1}(\theta) (\phi'(\theta) )^2 \right),
\end{eqnarray}
where $\mathcal I(\theta):= \mathbb E \left[ \left( \frac{\partial}{\partial \theta} \log \mathcal L (D | \theta) \right)^2 | \theta \right]$ is the Fisher information matrix, the expectation being taken with respect to the data vector $D.$ This last result allows to evaluate the standard error of the MLE and tells us that it decreases as $1/\sqrt{n}.$ 

However, note that the asymptotic variance in (\ref{eq:asymptotic}) is not directly available, since it depends on $\theta,$ which is unknown. In practice, it must be estimated, for instance by its MLE $\frac{1}{n}\mathcal I^{-1}(\hat \theta_\textsc{mle}) (\phi'(\hat \theta_\textsc{mle}) )^2 .$ Furthermore, these good large sample properties do not constitute a sufficient reason to choose MLE over Bayes estimates, because the latter also behave satisfyingly in presence of many data. Indeed, Doob's theorem ensures that, under mild conditions regarding the prior, the posterior distribution concentrates around the true parameter value \cite{Vandervaart00}. Furthermore, under essentially the same regularity assumptions as for the MLE, the posterior distribution of $\phi$ can be well approximated by :
\begin{eqnarray}\label{eq:vonMises}
\mathscr L (\phi | D) &\approx& \mathcal N \left(\widehat{\phi}_{\textsc{mle}} , \frac{1}{n}\mathcal I^{-1}(\hat \theta_\textsc{mle}) (\phi'(\hat \theta_\textsc{mle}) )^2 \right),
\end{eqnarray}
as shown for instance in \cite{Berger85}. Despite the striking similarity between (\ref{eq:vonMises}) and (\ref{eq:asymptotic}), it must be stressed that they have in fact fundamentally different meanings. Indeed, (\ref{eq:asymptotic}) describes the variability of $\widehat{\phi}_{\textsc{mle}}$ as a function of the random sample $D,$ conditional on the unknown value of the parameter $\theta.$ In contrast, (\ref{eq:vonMises}) refers to the posterior distribution of plausible values for the unknown quantity $\phi,$ conditional on the observed data $D.$

The main point of this comparison is that, when many observations are available, MLE and Bayes estimates become indistinguishable, whatever the choice of a prior and cost function, as the posterior variance of $\phi$ around $\widehat \phi_\textsc{mle}$ tends to zero. Thus, in this case there is no reason of preferring one approach over the other, though one may argue that the MLE has the advantage of simplicity, since it does not require specifying a cost function, nor a prior distribution.

\subsection{Small sample comparison}

In industrial studies however, more than often very few observations are at our disposal. This is especially the case when predicting the occurrence of some extreme event (such as the loss of a plane's steering mechanism), for which by definition very few observations have fortunately been made. In such cases the MLE looses all theoretical justification, and may perform very badly for a given loss function when compared to the Bayes estimate, which remains optimal in the sense described in Section~\ref{sec:Bayes}. The MLE may even be inadmissible, as in the classical example of estimating normal means in dimension greater than 2 \cite{Stein56}, or when estimating the shape parameter of a Weibull distribution \cite{Keller10}. In these cases, its use must thus be avoided.

Intuitively, the main reason for this poor behavior is due to the fact the MLE neglects the uncertainty on the parameter $\theta$ by assimilating it to its most likely value. However, especially for small values of $n,$ $\widehat{\phi}_{\textsc{mle}}$ may be unstable with respect to variations of the data, and take significantly different values from $\phi.$ This is precisely the issue addressed by the Bayesian approach, since instead of considering the loss in just one point, it averages it across all `probable' values of $\theta,$ given the information provided by both the prior and the data. This yields a more robust estimator (in the frequentist sense), the more so at small samples, since then the influence of the prior is non-negligible and tends to counterbalance sample variability, that is, variations in the observed variables $D.$

\section{Heuristic predictive approach}\label{sec:predictive}

\subsection{Principle}

\begin{figure}
\centering
\includegraphics[width=0.8\textwidth]{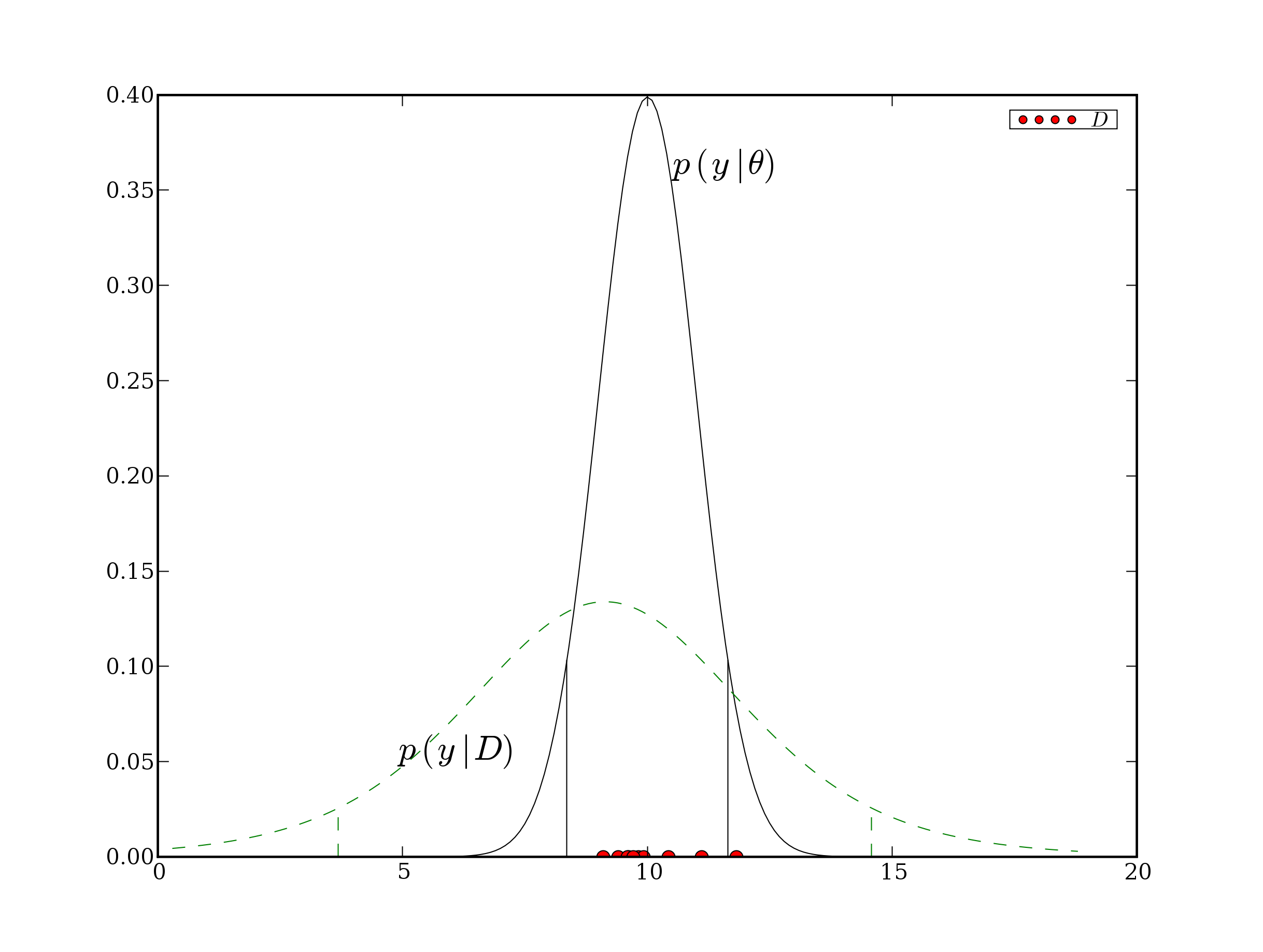}
\caption{\label{fig:normal_predictive} \small Density of the $\mathcal N(10, 1)$ distribution (continuous line) and corresponding posterior predictive distribution (dashed line), estimated from $n=10$ observations (dots), with a normal-inverse Gamma prior. Vertical segments correspond to the $5$-th and $95$-th percentiles of each density.}
\end{figure}

As we have seen in Section~\ref{sec:introduction}, a common way to account for the overall uncertainty on both observable and non observable quantities is to replace the unknown pdf $p(y | \theta)$ by the {\em posterior predictive density:}
\begin{eqnarray}\label{eq:predictive}
p(y | D) &:=& \int_{\theta} p(y | \theta) \pi(\theta | D) d \theta,
\end{eqnarray}
that is, to average the pdf over all possible values of $\theta,$ weighted by their probability given the information provided by the data $D.$ Here the output $\rm Y$ is interpreted as a future observation we wish to predict. For instance, in the survival analysis example introduced in Section~\ref{sec:decision}, $\rm Y$ could be seen as the lifetime of the ($n + 1$)-th industrial component, to be predicted based on the $n$ previously observed lifetimes. As illustrated in Figure~\ref{fig:normal_predictive}, the predictive density is typically more spread out than the true density, because uncertainty on $\theta$ has been added to the variability of $\rm Y.$

A seemingly natural use of the predictive density is to estimate any characteristic quantity of the (unknown) density $p(y | \theta),$ such as expected values, tail probabilities, quantiles, etc., by the corresponding characteristic quantity of $p(y | D)$ \cite{Geisser71,Christensen85}. Indeed, this provides a generic procedure to estimate any quantity of interest $\phi.$ Furthermore, this procedure has the advantage of accounting for the uncertainty on $\theta,$ having weighted each of its possible values according to their probability given the data. In the following, we will refer to this estimate as the {\em heuristic predictive estimator} (HPE).

Variants of the posterior predictive $p(y | D)$ are also commonly used, representing the uncertainty on $\theta$ by other densities than its posterior. These include:

\begin{itemize}

\item[--] The prior density $\pi(\theta)$ (see \cite{Helton96,Krzysztofowicz99} for instance). This yields the {\em prior predictive density}:
\begin{eqnarray}\label{eq:marginal}
p(y) &:=& \int_{\theta} p(y | \theta) \pi(\theta) d \theta,
\end{eqnarray}
which is thus defined independently from the data $D,$ or more precisely with $D = \{\},$ that is, in absence of data. In this case, inference on the output $\rm Y$ is purely based on prior, or expert, knowledge concerning the physical process generating the data. This is the case in \cite{Helton96}, where the future performance of a waste disposal plant is assessed based solely on the outputs of computer programs. These programs encode the experts' view of the plant's functioning and environment.

\item[--] The normal $\displaystyle p(\theta) = \mathcal N \left( \widehat{\theta}_{\textsc{mle}}, \frac{1}{n} \mathcal I^{-1}(\widehat{\theta}_{\textsc{mle}}) \right)$ as suggested in \cite{deRocquigny08}. From Section~\ref{sec:asymptotic}, this can be seen as the asymptotic approximation (\ref{eq:asymptotic}) of the sample distribution of $\hat \theta_\textsc{mle}$ (conditional on $\theta$), estimated by MLE.

However this interpretation is not coherent with the definition (\ref{eq:predictive}) of the predictive density, which requires the distribution of possible values of $\theta$ to be conditional on the data $D;$ a more appropriate interpretation would be to consider the above Gaussian as the asymptotic approximation (\ref{eq:vonMises}) to the posterior distribution $\pi(\theta | D).$ Thus it may be used as a convenient substitute when a large number of observations are available.

But, we have already stressed that large datasets are rather exceptional in real-life industrial studies. Hence for small samples, other types of posterior density approximations are required, such as  provided by stochastic sampling, {\em e.g.} Monte-Carlo Markov Chain (MCMC), techniques.

\end{itemize}

A practical implementation of HPE is the double Monte-Carlo algorithm. It consists in generating random values $\theta_1, \ldots, \theta_I$ from the posterior density $\pi(\theta | D)$ (or one of the above-mentioned alternatives), than simulating for each~$i = 1, \ldots, I$ one output $y_i$ from the conditional density $p(y_i | \theta_i).$ This generates a sample $y_1, \ldots, y_I$ from the predictive density (\ref{eq:predictive}), from which one can easily construct Monte-Carlo estimates of any desired quantity. Thus the double Monte-Carlo algorithm is essentially a technique to numerically integrate the joint posterior density of $(\theta, \rm Y).$

\subsection{Interpreting the predictive density}

As described above, the principle of HPE is to replace the sample density $p(y|\theta),$ representing the true physical phenomena under study, and available only in a perfect information setting, by the predictive density $p(y|D),$ which accounts for the uncertainty on $\theta$ when it is not known precisely. This use of $p(y|D)$ as a surrogate for $p(y|\theta)$ suggests that both densities are, to a certain point, exchangeable. The goal of this section is to stress that, though very similar in their mathematical expressions, these distributions have in fact very different meanings, and should not be mistaken for one another.

To see this, consider the following toy examples:

\begin{enumerate}

\item {\bf Coin tossing.} Suppose a well balanced coin is flipped 8 times, and each flip is recorded ($1$ for `heads', $0$ for `tails'). Any sequence of 8 binary digits is equally probable, so the result of this experience might be:
\begin{equation}\label{eq:coin_flip}
1\ 1\ 1\ 0\ 1\ 1\ 1\ 1.
\end{equation}
If asked what are the chances that the next throw will land on `heads', the obvious answer is $1/2,$ since we now that the coin is well-balanced. Note that this future throw is totally independent from the previously recorded ones.

\item {\bf Light bulbs, revisited.} Now assume light bulbs coming out from a production chain are tested, and their status reported ($1$ for a correctly working bulb, $0$ for a faulty one). Suppose the the $8$ first light bulbs lead to the following sequence:
$$
1\ 1\ 1\ 0\ 1\ 1\ 1\ 1.
$$
If asked what are the chances that the $9$-th light bulb will work correctly, a natural answer would be $7/8,$ based on the fact that $7$ out of the $8$ previous bulbs where working. Note that here the status of the future light bulb is considered dependent on that of the previously observed ones.

\end{enumerate}

Interpreting the above examples using the notations introduced in Section~\ref{sec:formulation}, in both cases the data $D,$ equal to (\ref{eq:coin_flip}), consists in $n=8$ independent realizations of a Bernoulli variable $\rm X,$ with $P[\rm X = 1|\theta] = \theta.$ The quantity of interest $\phi$ is the probability that a future realization $\rm Y$ of the same law is equal to $1,$ that is, $\phi(\theta) = P[\rm Y = 1|\theta] = \theta.$

The only difference between both examples is that in the first case $\theta$ is known. Hence, all statements are made conditional on $\theta,$ the future coin flip $\rm Y$ is independent from the previous ones. This is an example of perfect information, because the distribution $f(y|\theta)$ of the future observation is available; the uncertainty it describes is purely {\em aleatory}, in that it corresponds to the variability associated to a natural phenomenon (a coin toss).

In contrast, in example $2$ the proportion $\theta$ of well functioning light-bulbs is not known in advance, so it must be estimated from a sample $D$ of the production chain. Hence in that case the status $\rm Y$ of the future light bulb is not independent from the previously recorder ones. This is an example of imperfect information, the distribution of the future observation $f(y|\theta)$ being unavailable. The predictive distribution $f(y|D)$ can be used instead; in fact, $7/8$ is easily seen to be the HPE of $P[Y = 1|\theta],$ relative to the improper beta prior $B(0, 0).$

Note that the uncertainty described by $f(y|D)$ mixes the {\em aleatory} uncertainty associated to $\rm Y$ with the {\em epistemic} uncertainty on the parameter value. As such, it looses all phenomenologic interpretation. Rather, the predictive density represents one's current state of knowledge concerning the possible values of $\rm Y.$ After all, as stated in \cite{Goldstein10}, `Bayesian theory only describes how an individual's uncertainties are formed and modified by evidence'.

In conclusion, in both examples, the same terminology is used to talk about `the probability that the future observation $\rm Y$ equals 1'. However, in the first case, this refers to the true probability of a natural phenomenon, while in the second the probability is a subjective one, reflecting what the analyst is ready to bet on the outcome of an uncertain experience, in the sense of \cite{Savage54,deFinetti74}.

The risk of shortcuts in mixing epistemic and aleatory uncertainties extends in fact to all Bayesian analysis \cite{Helton94,Hofer96}, though it is most apparent in the predictive density. \cite{Ferson96} for instance advocates the use of interval analysis as an alternative to propagate epistemic uncertainty, combined with a probabilistic modeling of intrinsically random variables. Such a hybrid approach has the advantage of explicitly separating both sources of indetermination, and also avoids the issue of choosing a prior distribution. However it may raise other problems; for instance it is not clear wether the optimality properties of Bayesian estimates (see Section~\ref{sec:Bayes}) are retained in this setting, or how a combination of probability distribution and intervals can be used in a decision-making process. Hence, though we acknowledge the risk of confusion inherent to modeling parametric uncertainty by a probability distribution, we argue that it is still perfectly feasible to distinguish between both sources of uncertainty (see \cite{Helton96} for an example in the context of nuclear waste disposal).

In the following, we show that an important guideline for avoiding such issues is to focus on the decisional framework underlying the uncertainty analysis, while avoiding justifications based on heuristic interpretations.

\subsection{Estimation of an expected value (including failure probability)}

We now derive the HPE of the expected value of the output $\rm Y,$ that is, the integral of $\rm Y$ with respect to its density, so that the quantity of interest is:
\begin{eqnarray}
\phi = \mathbb E[{\rm Y} | \theta] &=& \int_y y\, p(y | \theta) dy.\nonumber
\end{eqnarray}
This is estimated by the {\em predictive mean}:
\begin{eqnarray}
\hat \phi_{\textsc{hpe}} = \mathbb E[{\rm Y} | D] &=& \int_y y\, p(y | D) dy.\nonumber
\end{eqnarray}
However, recalling that $p(y | D)$ is itself defined in (\ref{eq:predictive}) as an integral, and exchanging the order of integration,we can re-write the predictive mean as:
\begin{eqnarray}
\mathbb E[{\rm Y} | D] &=& \int_{\theta} \left\{ \int_y y\, p(y | \theta) dy \right\} \pi(\theta | D) d\theta = \int_{\theta} \phi (\theta) \pi(\theta | D) d\theta \nonumber\\
&=& \mathbb E[\phi | D].\nonumber
\end{eqnarray}
In other terms, in this case $\phi$ is simply estimated by its posterior mean! Furthermore, as recalled in Section~\ref{sec:examples}, the posterior mean is the Bayes estimate associated to the quadratic loss function. Thus, applying the HPE approach to $\mathbb E[{\rm Y} | \theta],$ we have actually performed a Bayesian estimation, implicitly assuming a quadratic loss, which may seem reasonable in this case.

A practical consequence of this result is that, if $\mathbb E[{\rm Y} | \theta]$ is available in closed-form, than its posterior mean can be evaluated by a simple Monte-Carlo algorithm, requiring only a sample from $\theta'$s posterior distribution. Hence, the double Monte-Carlo procedure described at the end of the previous section is unnecessarily complicated in this case.

The same reasoning applies in fact to any interest function $\phi$ that can be expressed as the expectation of a function of $\rm Y.$ This includes the expectation of any power of $\rm Y,$ or the probability that $\rm Y$ exceeds a certain value $t,$ which can be written as the expectation of $\bs 1_{\{ {\rm Y} > t \}}.$ The problem of estimating such a probability, arises in the context of structural reliability, where the failure of a system occurs when a given state variable is greater (or lower) than a fixed `safety limit' (see for instance the lifetime example presented in Section~\ref{sec:formulation}).

We may therefore state the following theorem (also found in \cite{Christensen85}), whose proof is given in Appendix~\ref{app:predictive_expectation}:
\begin{theorem}{ \bf Heuristic predictive estimate of an expected value.}\label{theo:predictive_expectation}

The HPE of any quantity of interest that can be written under the form $\phi = \mathbb E[h(y) | \theta]$ for a certain function $h$ is the posterior mean:
\begin{eqnarray}
\hat \phi_\textsc{hpe} &=& \mathbb E[\phi | D],\nonumber
\end{eqnarray}
that is, the Bayes estimate associated to the quadratic cost function. 

\end{theorem}

This result raises some concerns, since it suggests that the HPE is actually a Bayesian estimator in disguise. Moreover, this heuristic implicitly forces the choice of a particular cost function, which seems to depend on the expression of the quantity of interest, rather than on decisional aspects. For instance, as mentioned earlier in Section~\ref{sec:examples}, when estimating the probability that an industrial component's lifetime exceeds a certain length, under and over estimations may have very different consequences, so we may want to use a dissymmetric cost function in this case rather than the quadratic loss.

\subsection{Estimation of a quantile}

Can the HPE always be interpreted as a Bayes estimator? In the case of quantiles, there is no simple answer to that question.

To make this question more precise, consider $\phi$ to be the quantile $q_\alpha = q_\alpha (\theta)$ of order $\alpha$ of the density of $\rm Y,$ defined by:
\begin{eqnarray}
\mathbb P[{\rm Y} \leq q_\alpha | \theta] = \int_{-\infty}^{y = q_\alpha} p(y | \theta) dy &=& \alpha.
\end{eqnarray}
Then the HPE of $q_\alpha$ is the corresponding {\em predictive quantile} $\widehat q_\alpha^\textsc{hpe},$ such that:
\begin{eqnarray}
\mathbb P[{\rm Y} \leq \widehat q_\alpha^\textsc{hpe} | D] = \int_{-\infty}^{y = \widehat q_\alpha^\textsc{hpe}} p(y | D) dy &=& \alpha.\nonumber
\end{eqnarray}

A natural question that arises at this point is wether $\widehat q_\alpha^\textsc{hpe}$ is a Bayes estimate of $q_\alpha,$ that is, wether it can be defined as the quantity that minimizes the posterior expected loss $\mathbb E[C(q_\alpha, d) | D]$ (\ref{eq:expected_loss}), for a certain cost function $C(q_\alpha, d)$ (as defined in Section~\ref{sec:cost}). Intuitively, this would mean that the HPE of $q_\alpha,$ which is a characteristic quantity of the predictive density $p(y | D)$ of a future observation $\rm Y,$ would also be a certain characteristic quantity of the posterior density $\pi(q_\alpha | D)$ of quantile $q_\alpha,$ such as its posterior mean, standard deviation, etc. Unfortunately, such is not the case, as we prove in Appendix~\ref{app:predictive_quantile}, meaning that $\widehat q_\alpha^\textsc{hpe}$ is {\em not} a Bayes estimate of $q_\alpha.$

Then what exactly is $\widehat q_\alpha^\textsc{hpe}$? The answer is that it can be interpreted as a Bayes estimator, but of another quantity  than quantile $q_\alpha.$ Indeed, from Section~\ref{sec:examples}, $\widehat q_\alpha^\textsc{hpe}$ is seen to be formally the Bayes estimate of $\rm Y$ relative to the weighted absolute loss:
\begin{eqnarray}\label{eq:absolute_loss2}
c_\alpha(y, d) &=& |y - d| \times ( \alpha\bs 1_{\{d < y\}}  + (1 - \alpha)\bs 1_{\{d > y\}} ).
\end{eqnarray}
Note however that in the classical definition of a cost function, as recalled in Section~\ref{sec:cost}, the interest quantity is a function $\phi(\theta)$ of the model parameters, whereas here it is seen to be the value $y$ taken by the future observation $\rm Y.$ So $\widehat q_\alpha^\textsc{hpe}$ is a Bayes estimator in a more general sense than the one specified in Section~\ref{sec:Bayes}.

The above interpretation of $\widehat q_\alpha^\textsc{hpe}$ as a Bayes predictor of the future observation~$\rm Y$ is reasonable when $\rm Y$ is central to the decision process. Such a situation is described in \cite{Parent07}, in the context of dam conception. Here $\rm Y$ represents the yearly maximal water level of a river, and $d$ the height of a dam to be constructed next to the same river. Using the absolute weighted loss, the optimal dam height is seen to be precisely the HPE $\hat q_\alpha^\textsc{hpe},$ where $\alpha$ is the relative cost resulting from a flood (due to an undersized dam), when compared to the sum of the global costs (expected damages + sure investments).

However, if the decision-maker needs to consider costs that are directly related to the actual value of the quantile $q_\alpha,$ then the HPE should not be used, since it addresses a decisional problem that has nothing to do with quantiles. A simple approach in this case would be for instance to adopt the quadratic cost function (\ref{eq:quadratic_loss}), in which case the Bayesian estimate of $q_\alpha$ is its posterior mean.

Alternative choices are proposed in \cite{Hickey09}, in the context of ecological risk assessment, the quantity of interest being the minimal concentration $HC_p$ of a certain chemical hazardous to $p\%$ of the species in a given habitat. This is expressed as the $p$-th percentile of the species sensitivity distribution (SSD), representing the distribution of tolerance values to the target chemical for a randomly sampled species within the studied habitat. It is argued that the use of the weighted absolute loss or the LINEX loss allows to rationally choose an estimator that optimizes the costs associated with over- and under-estimations. The weighted absolute loss yields a certain quantile of the posterior distribution of $HC_p,$ whereas the LINEX leads to a more complicated estimator, which has no explicit form but can be evaluated numerically. Note that, in this context, the HPE would be totally useless to estimate $HC_p,$ since it would be unable to account for the costs resulting from the different types of estimation errors.

In conclusion, we summarize two important facts concerning the HPE of a quantile in the following theorem:

\begin{theorem}[\bf Heuristic predictive estimate of a quantile.]\label{theo:predictive_quantile}
$ $
\begin{enumerate}
\item The HPE $\widehat q_\alpha^\textsc{hpe}$ of a given quantile $q_\alpha$ is not a Bayes estimate of $q_\alpha$;
\item $\widehat q_\alpha^\textsc{hpe}$ is the Bayes estimate of a future observation $\rm Y$ for the weighted absolute loss $c_\alpha(y, d)$ (\ref{eq:absolute_loss2}).
\end{enumerate}

\end{theorem}

This is a striking illustration of how careful one must be when constructing an estimator in specifying both the quantity of interest and an expression, even imperfect, of the real costs associated to the decisions guided by the estimate, as stressed in \cite{Bernier03}. Indeed, neglecting these steps may lead to estimators that are totally unrelated to the quantity of interest from a decisional viewpoint, and may thus under- or over- estimate resulting costs in an unpredictable way.

\section{Example: Dyke reliability estimation}\label{sec:dyke}

We now compare the MLE, HPE  and Bayes estimates on a case study, concerning the safety evaluation of a flood protection dyke. There exists a rich literature on this subject, an overview of which can be found in \cite{Miquel84}. Following the terminology introduced in this work, the following example uses the {\em method of yearly maxima}, meaning that the flood probability is estimated from a record of yearly maximal river discharges.

The variable of interest $Y$ is here the maximal water level, noted $Z_c$ in the following, of the river at the location of the dyke. Following \cite{Pasanisi09}, we assume that $Z_c$ can be computed given a number of input variables, following the analytical formula:
\begin{eqnarray}\label{eq:hydrological_model}
Z_c &=& Z_v + A \cdot Q^{3/5},
\end{eqnarray}
where:
\begin{itemize}
\item[-] $Q$ is the yearly maximal water discharge ($\rm {m^3/s}$);
\item[-] $Z_v$ is the riverbed level (m asl) at the downstream part of the river section under investigation;
\item[-] $A$ is a certain constant, depending on the Strickler friction coefficient, as well as on the slope, width and length of the river section.
\end{itemize}
Additionally, we note $h$ the height of the protection dyke, and consider the problem of estimating the probability of a flood, that is, the probability that the maximal water level $Z_c$ exceeds $h.$ As in the light-bulb example discussed in Section~\ref{sec:formulation}, we assume that the estimated flood probability $\hat p$ will guide a certain decision. For instance, if $\hat p \leq 10^{-3},$ then the dyke is considered safe and no particular measure is taken. If $10^{-3} < \hat p \leq 10^{-2},$ then the valley downhill of the dyke is declared a flood-risk area, and constructions are forbidden therein. If $\hat p > 10^{-2},$ then enlarging the dyke is considered necessary.

Each decision $\hat p$ has a certain cost, depending on the real but unknown true flood probability $p,$ and which can be described by a cost function $C(p, \hat p),$ as explained in Section~\ref{sec:cost}. It is reasonable to assume for instance that the cost $C(p = 10^{-1}, \hat p = 10^{-4})$ resulting from under-estimating the flood probability, is more important than the false alarm costs $C(p = 10^{-3}, \hat p = 5.10^{-2}),$ corresponding to the scenario where new constructions are needlessly banned from the vicinity of the dyke, or even $C(p = 10^{-3}, \hat p = 10^{-1}),$ relative to the unnecessary efforts to enlarge the dyke.

The case presented here is of course an over-simplified toy example, meant for illustrative purposes only, and is not representative of the models used to assess hydrological risks in real-life industrial studies.

\subsection{Observation model and prior}

Further simplifying the case-study in \cite{Pasanisi09}, we assume that $Z_v$ and $A$ are known, fixed quantities. The yearly maximum river flow $Q$ on the other hand is intrinsically random, and we suppose that a sample of $n=30$ annual maximal values $D = (q_1, \ldots, q_n)$ is available, and can be used to assess the uncertainty on $Q.$ According to \cite{Miquel84}, $n=30$ is considered as the minimal sample size allowing a reliable estimate of the flood probability. For the exemplary purposes of this paper, we chose to model the annual maxima of the river discharge according to the Weibull distribution $\mathcal W(\eta, \beta),$ with scale parameter $\eta > 0$ and shape parameter $\beta > 0.$ This distribution has been proposed in the same context for instance in \cite{Feuillet87}, with an additional parameter $Q_0$ representing a minimal discharge. According to this model, the probability that $Q$ exceeds a certain level $t$ is given by:
\begin{eqnarray}
\mathbb P [Q > t | \eta, \beta] 
&=& \exp \left\{ -(t / \eta)^\beta \right\}.\nonumber
\end{eqnarray}
Given the simplified hydrological model~(\ref{eq:hydrological_model}) considered here, this means that the flood probability $p$ we are interested in is related simply to the parameters $(\eta, \beta),$ following:
\begin{eqnarray}\label{eq:dyke_reliability}
p &=& \mathbb P [Z_c > h | \eta, \beta] \nonumber\\
&=& \mathbb P [Z_v + A \cdot Q^{3/5} > h | \eta, \beta] \nonumber\\
&=& \exp \left\{ -\left(\frac {\left((h - Z_v)/ A\right)^{5/3}} \eta \right)^\beta\right\}.
\end{eqnarray}
Following \cite{Bousquet09}, we adopt a hierarchical form for the prior density,
by defining $\pi(\eta | \beta)$ as a generalized inverse Gamma (GIG) distribution, meaning that $\mu = \eta^{-1 / \beta}$ follows a Gamma distribution conditional on $\beta.$ We also choose a Gamma prior for $\beta,$ adding a lower bound to ensure existence of the posterior moments of $\eta$ \cite{Sun05}, so that:
\begin{eqnarray}
\pi(\mu | \beta) &=& \mathcal G \left( \mu ; m, b(m, \beta) \right)\nonumber\\
\pi(\beta) &\propto& \mathcal G \left( \beta ; m, m / \beta_0 \right) \bs 1_{\beta > \beta_l}.\nonumber
\end{eqnarray}
Here $b(m, \beta) = \frac{t_e^\beta}{2^{1/m} - 1},$ $t_e$ is a prior guess on the median annual flood, $\beta_0$ is a prior guess on $\beta,$ and $m$ is a virtual data size which measures the confidence in the prior information.

\subsection{Methods compared.}

We compared several estimates for the tail probability $p.$ First, we considered the commonly used MLE $\hat p_\textsc{mle},$ obtained by plugging the most likely parameter values $(\widehat \eta_\textsc{mle}, \widehat \beta_\textsc{mle})$ into their unknown value in (\ref{eq:dyke_reliability}), together with the HPE $\hat p_\textsc{hpe},$ which, by a direct application of Theorem ~\ref{theo:predictive_expectation}, is simply the posterior mean $\mathbb E[ p | D]$ .

But, as discussed in Section~\ref{sec:costs}, we are less interested in the precise value of $p$ than by its order of magnitude, that is, $-\log_{10} p.$ Thus, we computed the Bayes estimate $\hat p_\textsc{bay,1}$ relative to the log-quadratic loss~(\ref{eq:log_quadratic_loss}) (equivalently, $-\log_{10} \hat p_\textsc{bay,1}$ is the Bayes estimate of $-\log_{10} p$ using the quadratic loss~(\ref{eq:quadratic_loss}), that is, the posterior mean $\mathbb E [-\log_{10} p | D]$).

Finally, the above Bayes estimate equally penalizes over- and under-estimation of the log-tail probability, even though the risks associated to each type of error are very different. Indeed, under-estimating the probability of a flood can result in huge environmental, economic and human cost, compared to the safety measures discussed in Section~\ref{sec:dyke}. Thus we also considered the Bayes estimate $\hat p_\textsc{bay,2}$ relative the log-weighted absolute loss, meaning that $-\log_{10} \hat p_\textsc{bay,2}$ is the Bayes estimate relative to the weighted absolute loss~(\ref{eq:absolute_loss}). As recalled in Section~\ref{sec:examples}, this is simply the $\alpha$-th quantile of the posterior distribution of $-\log_{10} p,$ where $\alpha = C_1 / (C_1 + C_2)$ measures the relative cost of under-estimating the flood probability order of magnitude. We chose $\alpha=0.1,$ meaning that we considered that under-estimating the flood probability was 9 times as costly as over-estimating it.

\begin{figure}
\centering
\includegraphics[width=0.8\textwidth]{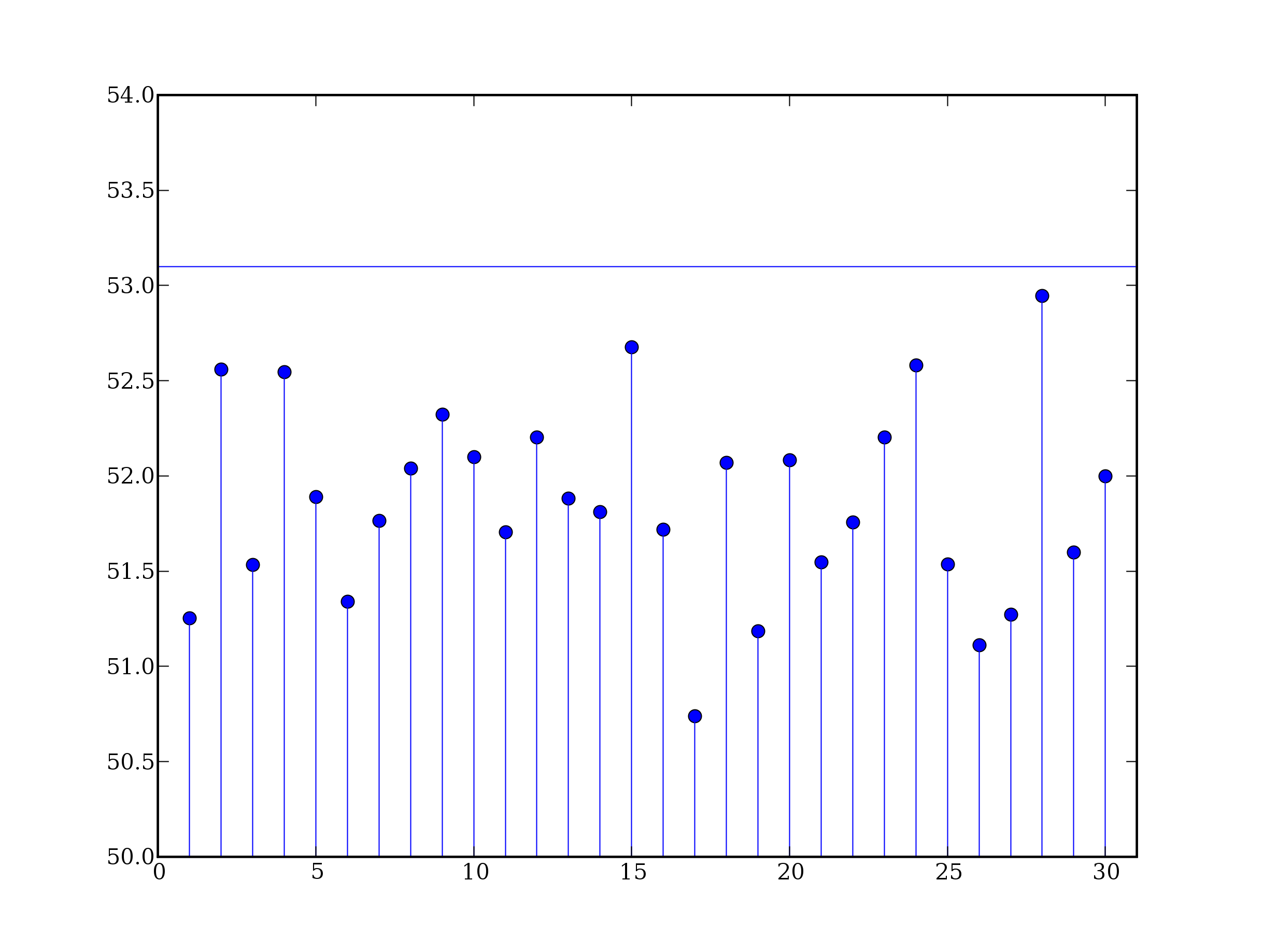}\\
\includegraphics[width=0.8\textwidth]{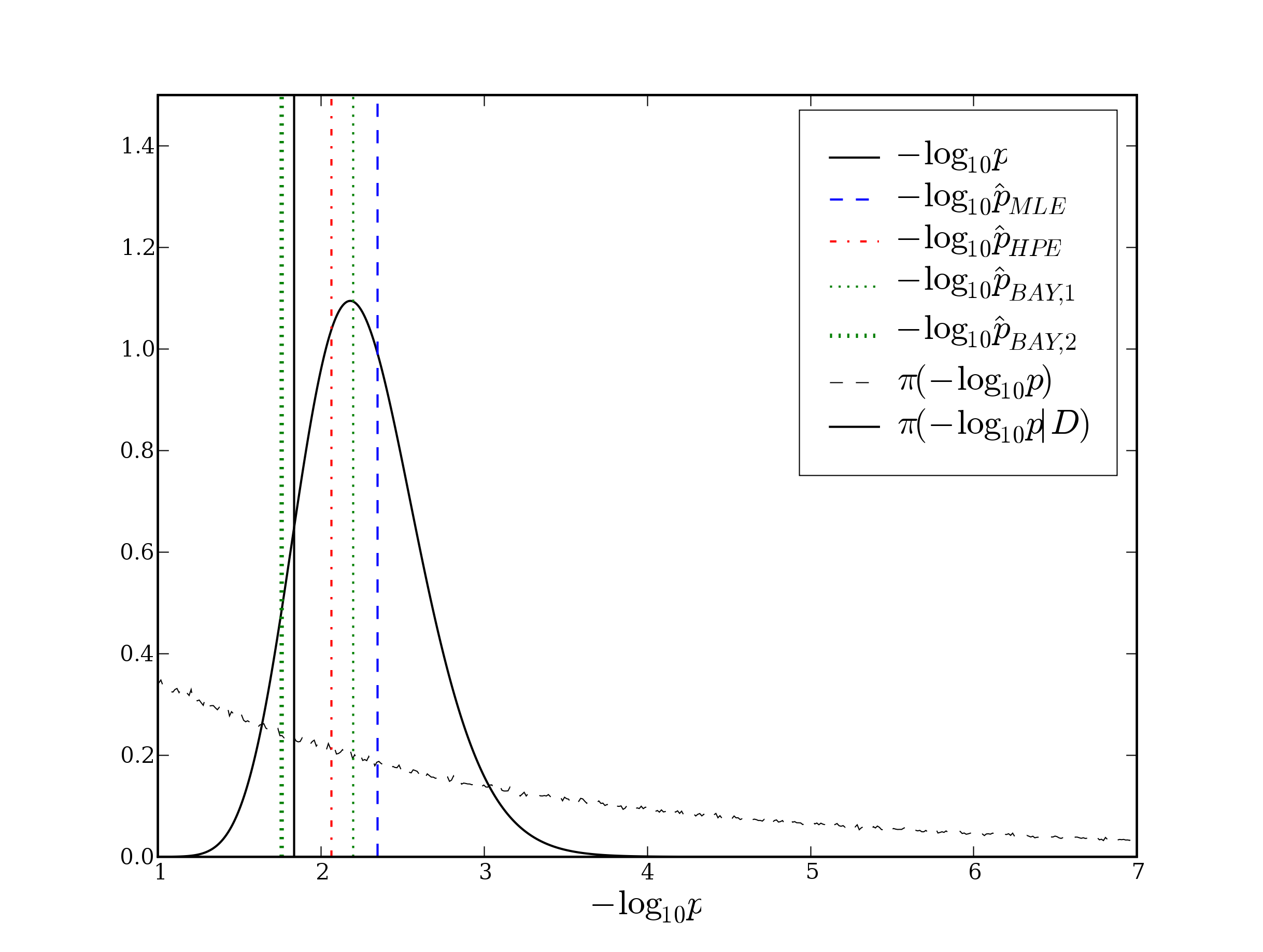}\\
\caption{\label{fig:dam_reliability_estimation} \small Top: $n=30$ maximal yearly water levels, computed using the hydrological model~(\ref{eq:hydrological_model}), from discharges simulated using the $\mathcal W (\eta=1\, 000, \beta=2)$ distribution. The horizontal line corresponds to the dyke height, $d=53.1\rm m.$ Bottom: Different estimates of a flood probability, on a logarithmic scale. The dyke height $d=53.1\rm m$ corresponds to a true flood probability of $p = .013.$ Prior bet on $\beta$ was taken equal to $\beta_0 = 1.5,$ and $t_e$ defined as the median of the Weibull distribution $\mathcal W(\beta_0, \eta_0),$ with prior scale $\eta_0 = 800.$ Limited weight was given to this prior information, by choosing $m = 1,$ explaining why the prior $\pi(-\log_{10} p)$ appears flat.}
\end{figure}

\subsection{Results.}

Figure~\ref{fig:dam_reliability_estimation} shows the values of the above estimators, when applied to a simulated dataset. The MLE was evaluated numerically using a standard Newton-Raphson algorithm to maximize the likelihood function, while a sample of $10^6$ draws from the parameters' joint posterior distribution was obtained using importance sampling (IS), and used to compute Monte-Carlo approximations of all Bayesian estimates.

In this particular example, the MLE $(\hat p_\textsc{mle} = .004),$ the HPE $(\hat p_\textsc{bay,2} = .009)$ and the Bayes estimate relative to the log-quadratic loss $(\hat p_\textsc{bay,1} = .006)$ all estimated the flood risk to be smaller than $10^{-2},$ even though the true probability was: $p = .013.$ According to the decisional setting described previously, based on these estimates the valley downhill from the dyke would be declared unconstructible. However, the danger of a flood occurring in less than $100$ years would go unforeseen.

In contrast, only the Bayes estimate relative to the log-weighted absolute loss $(\hat p_\textsc{bay,2} = .017)$ correctly evaluated the flood probability as exceeding $10^{-2},$ suggesting the necessity of re-sizing the dyke.

Hence the safest estimate (from a decisional point of view) in this case was $\hat p_\textsc{bay,2}.$ This is hardly surprising, since it was the only one based on a loss function that explicitly favored overestimation of the quantity of interest, through precise quantification of the relative cost resulting from the dyke being flooded, when compared to the establishment of safety measures, {\em i.e.}, enlargement of the dyke, or evacuation of the houses in the downhill valley. Because these costs are very different, it is important to consider a method of estimation which takes them into account.

\section{Discussion}\label{sec:discussion}

We have compared the main paradigms used to perform statistical inference in industrial studies, namely Bayesian inference, maximum likelihood estimation (MLE) and heuristic predictive estimation (HPE). This comparison was conducted with reference to decision theory, since industrial studies usually guide a certain decision. The Bayes paradigm is known to be optimal from this perspective, hence we used it as a gold standard to benchmark both MLE and HPE.

MLE is by far the most popular approach due to its simplicity, however it is justified only when a large number of observations are available. Its main weakness is that it ignores the uncertainty on the unknown parameter, making it unreliable when only a few observations are available.

HPE represents a definite improvement over the MLE in that it explicitly accounts for parameter uncertainty by averaging the output's density over all possible values of the parameter, weighted by their posterior probabilities. However, we have shown that, for a certain number of interest quantities, HPE is in fact equivalent to Bayes estimation, and corresponds implicitly to a certain choice of cost function, depending entirely on the mathematical expression of the chosen quantity of interest.

This raises several concerns. First, why use the term `predictive estimation' when actually Bayes estimation is performed? Second, cost functions should be chosen based on an evaluation of the costs associated with potential estimation errors, rather than the particular expression of the quantity under study.

In other terms, the major drawback of HPE is that it robs the user of the freedom of specifying exactly what estimation problem he wishes to solve, by implicitly choosing for him the cost function to be minimized. Furthermore, we have seen that interpreting the results of HPE is delicate when one steps outside the realm of Bayesian rationale, since the predictive density has no phenomenological interpretation, and should only be interpreted in terms of probabilistic bets.

In conclusion, in view of the increasing need to assess uncertainties in industrial studies, and the ever more pressing requirement of accurate cost assessment, the Bayesian framework seems to provide the most useful unified framework. Indeed, it relieves the engineer from the difficult choice of an appropriate methodology, allowing him/her to concentrate on the really important questions underlying the study, which are the quantification of the sources of uncertainties, and that of the potential costs to be controlled in the decision-making process. Furthermore, adopting such a unified framework would make it much simpler to compare results from different studies, an important issue stressed in \cite{Pate-cornell96}.

It does imply some conceptual and technical difficulties, such as choosing a prior distribution, or computing the joint posterior distribution of all unknown variables, which may explain why the Bayesian paradigm has not yet been universally adopted. However, many solutions exist nowadays to tackle these issues. See \cite{Marin07} for instance for an excellent overview of the current state of the art in Bayesian estimation techniques, and \cite{Boreux10} for their application to engineering case studies. When comprehensive Bayesian approaches are unpractical, it is also possible to adopt simplifying strategies, such as in \cite{Krzysztofowicz99,Maranzano08}. These consist in focusing the Bayesian treatment on the most important sources of uncertainty with respect to the decisional problem considered, treating other sources by more conventional methods, such as MLE. Hence we have every reason to believe that more and more industrial studies will make use of the powerful tools of decision theory and Bayesian analysis \cite{Berger85} to conduct as simply and efficiently as possible the increasingly complex studies required by an increasingly complex world.

\section{Acknowledgements}
The authors gratefully thank Prof. Jacques Bernier and Dr. Nicolas Bousquet for their helpful advises. This work was partially supported by the French Ministry of Economy in the context of the CSDL (Complex Systems Design Lab) project of the Business Cluster ``System@tic Paris-R\'egion''.

\appendix

{
\small


}

\section{Proof of Theorem~\ref{theo:predictive_expectation}}\label{app:predictive_expectation}

By assumption, there exists a function $h,$ such that the quantity of interest $\phi$ can be written as:
\begin{eqnarray}
\phi = \mathbb E[h({\rm Y}) | \theta] = \int_y h(y) p(y | \bs  \theta) dy.\nonumber
\end{eqnarray}
This is estimated using the predictive heuristic by substituting $p(y | D)$ for the unknown density $p(y | \theta):$
\begin{eqnarray}\label{eq:predictive_expectation}
\hat \phi_\textsc{hpe} &=& \int_y h(y) p(y | D) dy.
\end{eqnarray}
Expanding the expression of $p(y | D),$ as defined in (\ref{eq:predictive}), and exchanging the order of integration then yields:

\begin{eqnarray}
\hat \phi_\textsc{hpe} &=& \int_{\theta} \left\{ h(y) \int_y p(y | \theta) dy \right\} \pi(\theta | D) d \theta = \int_{\theta} \phi (\theta ) \pi(\theta | D) d \theta \nonumber\\
&=& \mathbb E[\phi | D]\nonumber
\end{eqnarray}
\begin{flushright}
$\square$
\end{flushright}

\section{Proof of Theorem~\ref{theo:predictive_quantile}}\label{app:predictive_quantile}

Reasoning by {\em reductio ad absurdum}, suppose that $\widehat q_\alpha^\textsc{hpe}$ {\em is} the Bayes estimate of $q_\alpha,$ relative to a certain cost function~$C(q_\alpha, d),$ that is, the quantity that minimizes $\mathbb E[C(q_\alpha, d) | D],$ which we can write formally as:
\begin{eqnarray}
\widehat q_\alpha^\textsc{hpe} &=& \arg \min_d \int_z C(z, d) \pi_{q_\alpha | D}(z) d z,\nonumber
\end{eqnarray}
where we note $\pi_{q_\alpha | D}(z)$ the posterior density of quantile $q_\alpha,$ given data $D$ and prior $\pi.$ The right member of the above display defines a functional $\Xi$ whose domain is the space of all possible posterior densities of $q_\alpha,$ that is, the space of all probability densities $h(\cdot)$ on $\mathbb R.$ When $h$ coincides with the posterior density $\pi_{q_\alpha| D}$ of a certain quantile $q_\alpha,$ then we have: $\Xi(\pi_{q_\alpha| D}) = \widehat q_\alpha^\textsc{hpe}.$

We now show that the existence of such a functional leads to a contradiction in the exponential model. Using the notations introduced in the example in Section~\ref{sec:formulation}, the quantile $q_\alpha$ is given by:
\begin{eqnarray}
q_\alpha &=& \theta \log \frac{1}{1 - \alpha}.
\end{eqnarray}

Defining the usual conjugate inverse-Gamma prior on $\theta:$
\begin{eqnarray}
\pi(\theta) &=& \mathcal{IG} (\theta ; n_0, S_0)\\\nonumber
&=& \frac {S_0^{n_0}} {\Gamma(n_0)} \theta^{- n0 - 1} e^{-S_0 / \theta},
\end{eqnarray}
it is well known that the posterior distribution of $\theta$ is the inverse-Gamma distribution $\mathcal{IG}(n_0 + n, S_0 + S_n),$ where $n$ is the data size and $S_n$ the sum of all observations: $S_n = x_1 + \ldots + x_n.$ Hence, the posterior distribution of $q_\alpha$ is also an inverse-Gamma:
\begin{eqnarray}\label{eq:quantile_posterior}
q_\alpha | D &\sim& \mathcal{IG} \left(n_0 + n, \log \left( \frac{1}{1 - \alpha} \right) \times (S_0 + S_n)  \right) ,
\end{eqnarray}
and elementary calculations show that the HPE is equal to:
\begin{eqnarray}\label{eq:quantile_HPE}
\widehat q_\alpha^\textsc{hpe} &=& \left( e^{\log ( \frac{1}{1 - \alpha}) / (n_0 + n)} - 1 \right) \times (S_0 + S_n).
\end{eqnarray}

Note that, by definition of the functional $\Xi,$ the right member of (\ref{eq:quantile_HPE}) is the image by $\Xi$ of the posterior density defined by (\ref{eq:quantile_posterior}). The proof proceeds from here by showing that $\log (\frac{1}{1 - \alpha})$ itself is the image by a certain functional of the posterior density (\ref{eq:quantile_posterior}), {\em i.e.}, that one can recover the order $\alpha$ of a quantile from the quantile's posterior distribution only! This is of course absurd since we can see from (\ref{eq:quantile_posterior}) that the same posterior scale parameter value $\log \left( \frac{1}{1 - \alpha} \right) \times (S_0 + S_n)$ can derive from infinitely many possible couples of values for $\alpha$ and $(S_0 + S_n),$ meaning that two quantiles of different orders may have the exact same posterior distribution if computed on two different datasets, or for two different values of the prior scale $S_0.$

But it is easy to show that $\log (\frac{1}{1 - \alpha})$ can be deduced from $\widehat q_\alpha^\textsc{hpe},$ $(n_0 + n)$ and $\log \left( \frac{1}{1 - \alpha} \right) \times (S_0 + S_n).$ Since the shape and scale parameters of an inverse-Gamma distribution can be expressed as simple functions of its mean and variance, $(n_0 + n)$ and $\log \left( \frac{1}{1 - \alpha} \right) \times (S_0 + S_n)$ can also be written as the images by certain functionals of the posterior density (\ref{eq:quantile_posterior}). As a result, $\alpha$ can be obtained from (\ref{eq:quantile_posterior}) only, which we have shown to be impossible
\begin{flushright}
$\square$
\end{flushright}

\end{document}